\documentclass[final]{elsart}
\usepackage{ifpdf,graphicx,epsfig,bm,amsfonts,amssymb,amsmath}
\begin{document}

\begin{frontmatter}

\title{Conformally flat Kaluza-Klein spaces, \\ pseudo-/para-complex space forms \\
and generalized gravitational kinks}

\author[it]{Paolo~Maraner} and
%\ead{PMaraner@unibz.it}
\author[uk]{Jiannis~K.~Pachos}%\corauthref{cor}}
%\corauth[cor]{Corresponding author.}
%\ead{J.K.Pachos@leeds.ac.uk}

\address[it]{School of Economics and Management, Free University of Bozen-Bolzano,\\
             via Sernesi~1, 39100~Bolzano, Italy}
\address[uk]{School of Physics and Astronomy, University of Leeds, Leeds LS2 9JT, UK}

\begin{abstract}
The equations describing the Kaluza-Klein reduction of conformally flat
spaces are investigated in arbitrary dimensions. Special classes of solution
related to pseudo-K\"ahler and para-K\"ahler structures are constructed and
classified according to spacetime dimension, signature and gauge field rank.
Remarkably, rank two solutions include gravitational kinks together with
their centripetal and centrifugal deformations.
\end{abstract}


\end{frontmatter}

\section{Introduction}
\label{Intro}

In a recent paper \cite{Grumiller&Jackiw06} Grumiller and Jackiw
investigated the Kaluza-Klein reduction of conformally flat spaces from
$d+1$ to $d$ dimensions, for $d\geq3$. After obtaining appropriate reduction
formulas in terms of Kaluza-Klein functions, they imposed the vanishing of
the higher dimensional conformal tensor, producing equations  describing the
`immersion' of a codimension one spacetime into a conformally flat space.
Let us parameterize the higher dimensional line element as
$ds_{(d+1)}^2=g_{\mu\nu}dx^\mu dx^\nu +(A_\mu dx^\mu+ dx^{d})^2$, with Greek
indices ranging over $0,1, ..., d-1$ and all quantities independent of
the last coordinate $x^d$. Then, the Grumiller-Jackiw equations ((17a,b,c)
in Ref.~\cite{Grumiller&Jackiw06}) read
\begin{subequations}
\begin{eqnarray}
&&C_{\mu\nu\kappa\lambda} +
\frac{1}{2}\left(F_{\mu\nu}F_{\kappa\lambda}-
F_{\mu[\kappa}F_{\lambda]\nu}\right)
-\frac{3}{2(d-2)}\left(g_{\mu[\kappa} T_{\lambda]\nu}-
g_{\nu[\kappa}T_{\lambda]\mu}\right)=0 \label{GJ1},\\
&&R_{\mu\nu}-\frac{1}{d} R g_{\mu\nu}=\frac{d+1}{4}
\left(F_{\mu\kappa}{F_\nu}^\kappa
-\frac{1}{d}F^2g_{\mu\nu}\right)\label{GJ2},\\
&&D_\kappa F_{\mu\nu}+ \frac{2}{d-1}g_{\kappa[\mu}D_\lambda
{F_{\nu]}}^\lambda=0, \label{GJ3}
\end{eqnarray}\label{GJ}\end{subequations}
with $g_{\mu\nu}$ the $d$-dimensional spacetime metric, $D_\kappa$ the
associated covariant derivative, $C_{\mu\nu\kappa\lambda}$, $R_{\mu\nu}$,
$R$ the corresponding Weyl, Ricci and scalar curvatures,\footnote{Our
curvature conventions are
${R_{\mu\nu\kappa}}^\lambda=\partial_\mu\Gamma_{\nu\kappa}^\lambda-...$,
$R_{\mu\nu}={R_{\kappa\mu\nu}}^\kappa$, and $R={R_\mu}^\mu$.}
$F_{\mu\nu}=\partial_\mu A_\nu-\partial_\mu A_\nu$ the Kaluza-Klein  gauge
field, $F^2=F_{\mu\nu}F^{\mu\nu}$ its squared modulo,
$T_{\mu\nu}=F_{\mu\kappa}{F_\nu}^\kappa-\frac{1}{2(d-1)}F^2g_{\mu\nu}$ and
square brackets denoting antisymmetrization,
$t_{[\mu\nu]}=(t_{\mu\nu}-t_{\nu\mu})/2$. The spacetime metric $g_{\mu\nu}$
is here allowed to carry arbitrary signature, while the signature of the
extra coordinate $x^d$ is chosen, for definiteness, as positive. The case
with $x^d$ carrying a negative signature is straightforwardly obtained by
replacing the $d+1$-dimensional metric by its opposite and correspondingly
changing the sign of all scalar and sectional
curvatures.\\
 After addressing dimensional reduction for arbitrary
dimensions Grumiller and Jackiw specialized to $d=3$ and constructed special
solutions based on a further Ansatz of the three-dimensional metric. In this
note we investigate equations (\ref{GJ}a), (\ref{GJ}b) and (\ref{GJ}c) in
their full generality. We construct classes of solutions classified by
spacetime dimension, signature and by the rank of the Kaluza-Klein gauge
field. All solutions with non-vanishing gauge curvature are related to
pseudo-K\"ahler or para-K\"ahler structures. Of particular interest is the
case of rank two gauge fields, where exceptional kink solutions together
with their centripetal and centrifugal deformations
appear.\\
Our discussion proceeds as follows. In~\S\ref{RRScurvatures} we obtain
explicit expressions for the spacetime Riemann, Ricci and scalar curvatures
in terms of the metric, $g_{\mu\nu}$, and the gauge field, $F_{\mu\nu}$.
This allows us to write down, in
\S\ref{ICond}, integrability conditions providing the higher dimensional
generalization of the `gravitational kink' equations obtained by Guralnik,
Iorio, Jackiw and Pi from the Kaluza-Klein reduction of the gravitational
Chern-Simons term
\cite{Guralnik&Iorio&Jackiw&Pi03}. These equations are somehow
easier to solve than the original ones. Null, maximal and intermediate rank
solutions are eventually obtained in
\S\ref{NullMax} and \S\ref{Int} and their relation to
pseudo-K\"ahler and para-K\"ahler structures is discussed. Our conclusions
and a list of the obtained solutions are presented in
\S\ref{conclusions}.

\section{Riemann, Ricci and scalar curvatures}
\label{RRScurvatures}

Here we shall demonstrate that equations (\ref{GJ}a), (\ref{GJ}b) and
(\ref{GJ}c) allow to express the spacetime Riemann, Ricci and scalar
curvatures entirely in terms of $g_{\mu\nu}$ and $F_{\mu\nu}$, up to an
arbitrary constant. Equation (\ref{GJ1}) is solved by
\begin{equation}
R_{\mu\nu\kappa\lambda}=\texttt{r}_{\mu\nu\kappa\lambda}-\frac{1}{2}\left(F_{\mu\nu}F_{\kappa\lambda}-
F_{\mu[\kappa}F_{\lambda]\nu}\right),\label{Rmnkl}
\end{equation}
with $\texttt{r}_{\mu\nu\kappa\lambda}$ a tensor sharing the symmetries of
the Riemann tensor|not the Bianchi identities|satisfying the conditions
\begin{equation}
\texttt{r}_{\mu\nu\kappa\lambda}
+\frac{2}{d-2}\left(g_{\mu[\kappa}\texttt{r}_{\lambda]\nu}
-g_{\nu[\kappa}\texttt{r}_{\lambda]\mu}\right)-
\frac{2}{(d-1)(d-2)}\texttt{r}g_{\mu[\kappa}g_{\lambda]\nu}=0,
\end{equation}
with $\texttt{r}_{\mu\nu}={\texttt{r}_{\kappa\mu\nu}}^\kappa$ and
$\texttt{r}={\texttt{r}_\mu}^\mu$. These are $\frac{1}{12}(d+1)(d+2)(d-3)$
simultaneous linear equations in $\frac{1}{12}d^2(d^2-1)$ variables with
coefficients only depending on the spacetime metric, $g_{\mu\nu}$. The
general solution depends on $\frac{1}{2}d(d+1)$ parameters that are
functions of the coordinates and is obtained as
\begin{equation}
\texttt{r}_{\mu\nu\kappa\lambda}=2\left(g_{\mu[\lambda}\rho_{\kappa]\nu}-
g_{\nu[\lambda}\rho_{\kappa]\mu}\right)+2\rho
g_{\mu[\lambda}g_{\kappa]\nu}\label{rmn},
\end{equation}
with  $\rho_{\mu\nu}$ a traceless symmetric tensor and $\rho$ a
scalar.  The tensor $\rho_{\mu\nu}$ is determined by equation
(\ref{GJ2}). From (\ref{Rmnkl}) and  (\ref{rmn}) we have
$R_{\mu\nu}=(d-1)\rho g_{\mu\nu}+(d-2)\rho_{\mu\nu}+\frac{3}{4}
F_{\mu\kappa}{F_\nu}^\kappa$ and $R=d(d-1)\rho+\frac{3}{4}F^2$,
which substituted in (\ref{GJ2}) yield
\begin{equation}
\rho_{\mu\nu}=\frac{1}{4} \left(F_{\mu\kappa}{F_\nu}^\kappa
-\frac{1}{d}F^2g_{\mu\nu}\right).\label{rhomn}
\end{equation}
Eventually, the scalar $\rho$ is fixed by the contracted Bianchi identities
and equation (\ref{GJ3}). By inserting ($\ref{rhomn}$) in the above
expressions for the Ricci and scalar curvatures we obtain from $D_\nu
{R_\mu}^\nu=\frac{1}{2}D_\mu R$ the equation
\begin{equation}
(d-1)(d-2)D_\mu\rho=\frac{d+1}{2}D_\nu
F_{\mu\kappa}F^{\nu\kappa}-\frac{5d-4}{4d}D_\mu
F^2.\label{CBianchi}
\end{equation}
Contracting (\ref{GJ3}) with $F^{\kappa\nu}$ and by means of the gauge
theoretical Bianchi identities we also obtain $D_\nu
F_{\mu\kappa}F^{\nu\kappa}=\frac{d}{4}D_\mu F^2$, showing that the right
hand side of (\ref{CBianchi}) is indeed a total derivative. Integration
gives
\begin{equation}
\rho=\frac{d+4}{8d}F^2+k,\label{rho}
\end{equation}
where $k$ is a constant. Next, we substitute (\ref{rho}) and (\ref{rhomn})
in (\ref{rmn}). By employing this result and by successive contractions of
Eq. (\ref{Rmnkl}) we obtain the Riemann, Ricci and scalar spacetime
curvatures in terms of  the metric $g_{\mu\nu}$, the gauge field
$F_{\mu\nu}$ and the arbitrary constant $k$ as
\begin{subequations}\label{curvatures}\begin{eqnarray}
&&R_{\mu\nu\kappa\lambda}=2\left(k+\frac{1}{8}F^2\right)g_{\mu[\lambda}g_{\kappa]\nu}\nonumber\\
&&\hskip1.5cm-\frac{1}{2}\left(g_{\mu[\kappa}F_{\lambda]\xi}{F_\nu}^\xi-g_{\nu[\kappa}F_{\lambda]\xi}
{F_\mu}^\xi\right) -\frac{1}{2}\left(F_{\mu\nu}F_{\kappa\lambda}-
F_{\mu[\kappa}F_{\lambda]\nu}\right),\label{Riemann}\\
&& R_{\mu\nu}=(d-1)kg_{\mu\nu}+\frac{(d+1)}{8}F^2g_{\mu\nu}
+\frac{(d+1)}{4}F_{\mu\kappa}{F_\nu}^\kappa,\label{Ricci}\\
&&R=d(d-1)k+\frac{(d+1)(d+2)}{8}F^2.\label{scalar}
\end{eqnarray}\end{subequations}
Direct computation shows that the Riemann tensor (\ref{Riemann}) satisfies
the Bianchi integrability conditions $D_\xi R_{\mu\nu\kappa\lambda}+D_\nu
R_{\xi\mu\kappa\lambda}+D_\mu R_{\nu\xi\kappa\lambda}=0$, provided that
(\ref{GJ3}) is satisfied. The integration of Grumiller-Jackiw equations is,
therefore, reduced to the integration of (\ref{GJ3}) subject to
(\ref{curvatures}).

\section{Integrability conditions}
\label{ICond}

It is useful to establish integrability conditions for (\ref{GJ3})
subject to (\ref{curvatures}). Consider the covariant derivative
of (\ref{GJ3})
\begin{equation}
D_\lambda D_\kappa F_{\mu\nu}+
\frac{2}{d-1}g_{\kappa[\mu}D_\lambda D_\xi
{F_{\nu]}}^\xi=0.\label{NablaGJ3}
\end{equation}
Antisymmetrizing (\ref{NablaGJ3}) in $\kappa,\lambda$, reexpressing the
commutator of covariant derivatives in terms of the Riemann tensor and
inserting (\ref{curvatures}), we obtain
\begin{equation}
\frac{1}{d-1}D_\mu D_\kappa{F_\nu}^\kappa-\left(k+
\frac{1}{8}F^2\right)F_{\mu\nu}+
\frac{1}{4}{F_\mu}^\kappa{F_\kappa}^\lambda F_{\lambda\nu}=0.
\label{Int1}
\end{equation}
Symmetrizing this expression in $\mu,\nu$ we have $D_\mu
D_\kappa{F_\nu}^\kappa+ D_\nu D_\kappa{F_\mu}^\kappa=0$, showing that
\begin{equation}
K_\mu=\frac{1}{d-1}D_\nu {F_\mu}^\nu \label{Killing},
\end{equation}
is a Killing vector of our geometry, when it is not identically vanishing.
The existence of such a Killing was recognized by Grumiller and Jackiw in
the special case $d=3$ ((26b) in Ref.~\cite{Grumiller&Jackiw06}).
Contracting now (\ref{NablaGJ3}) with $g^{\kappa\lambda}$ and by means of
$D_\mu K_\nu+D_\nu K_\mu=0$, we obtain
\begin{equation}
\frac{1}{d-1}D_\mu D_\kappa{F_\nu}^\kappa+ \frac{1}{2}D^2
F_{\mu\nu} =0.
\label{traceless}
\end{equation}
When substituted in (\ref{Int1}), this yields the integrability conditions in the
form
\begin{equation}
\frac{1}{2}D^2 {F_\mu}^\nu+\left(k+
\frac{1}{8}F^2\right){F_\mu}^\nu-
\frac{1}{4}{F_\mu}^\kappa{F_\kappa}^\lambda{F_\lambda}^\nu=0.
\label{kinkEq}
\end{equation}
Equations (\ref{traceless})  and (\ref{kinkEq}) are the higher
dimensional analogue of the `traceless' and `gravitational kink'
equations obtained from the Kaluza-Klein reduction of the
gravitational Chern-Simons term \cite{Guralnik&Iorio&Jackiw&Pi03}.

\section{Null and maximal rank solutions}
\label{NullMax}

Equations (\ref{GJ3}) and (\ref{kinkEq}) are trivially solved by a vanishing
gauge curvature. The Riemann tensor (\ref{Riemann}) consequently reduces to
\begin{eqnarray}
&&R_{\mu\nu\kappa\lambda}=k
\left(g_{\mu\lambda}g_{\kappa\nu}-g_{\mu\kappa}g_{\lambda\nu}
\right),\label{const.sec.curv.}
\end{eqnarray}
revealing that spacetime is a real pseudo-Riemannian manifold with constant
sectional curvature $k$. When complete, spacetime is then a
\emph{real space form}, isomorphic to the pseudo-Euclidean real space
$\mathbb{R}^d_s$ for vanishing sectional curvature, to the real
pseudo-projective space $\mathbb{R}P^d_s$|or pseudo-sphere $S^d_s$|for
positive sectional curvature or to the real pseudo-hyperbolic space
$\mathbb{R}H^d_s$ for negative sectional curvature (see e.g. \S8 of
Ref.~\cite{O'Neill83}). The signature is arbitrary, $s=0,...,d$. For
Euclidean signature, $s=0$, these are the standard Euclidean space
$\mathbb{R}^d\equiv\mathbb{R}^d_0$, sphere $S^d\equiv S^d_0$ and hyperbolic
space $H^d\equiv \mathbb{R}H^d_0$. For Lorentzian signature, $s=1$, one
obtains the Minkowski $\textrm{M}_d\equiv\mathbb{R}^d_1$, deSitter
$\textrm{dS}_d\equiv S^d_1$ and anti-deSitter $\textrm{AdS}_d\equiv
\mathbb{R}H^d_1$ spacetimes, respectively. Summarizing we obtain
\begin{equation}
 \mathbb{R}^d_s(0)
  \hskip0.3cm\mbox{for}\hskip0.2cm k=0,\hskip0.3cm
 \mathbb{R}P^d_s(k)
  \hskip0.3cm\mbox{for}\hskip0.2cm k>0,\hskip0.3cm
 \mathbb{R}H^d_s(k)
  \hskip0.3cm\mbox{for}\hskip0.2cm k<0,
\end{equation}
where we denote in brackets the sectional curvature, $k$. Real space forms
are conformally flat themselves, so that metric and vector potential can be
conveniently displayed in the form
\begin{eqnarray}
&&g_{\mu\nu}=\frac{1}{\left(1+\frac{k}{4}
\eta_{\kappa\lambda}x^\kappa x^\lambda
\right)^2}\eta_{\mu\nu},\hskip0,5cm A_\mu=0,
\label{g,A_const.sec.curv.}
\end{eqnarray}
with $\eta_{\mu\nu}$ a pseudo-Euclidean metric carrying arbitrary
signature.

Besides null rank solutions, a second class of solutions can be obtained
when the Kaluza-Klein two-form, $F_{\mu\nu}$, has maximal rank,
$\mbox{rank}\{F_{\mu\nu}\}=d$. Given the antisymmetry of $F_{\mu\nu}$, this
is only possible in an even number of dimensions, $d=2\textrm{d}$. Equation
(\ref{GJ3}) is in fact trivially satisfied by a covariantly constant gauge
curvature
\begin{equation}
D_\kappa F_{\mu\nu}=0,
 \label{DF=0}
\end{equation}
a condition which is fully equivalent to the constancy of the
scalar $F^2$ or to the vanishing of the Killing vector $K^\mu$.
The integrability conditions (\ref{kinkEq}) consequently reduce to
\begin{equation}
\left(k+ \frac{1}{8}F^2\right){F_\mu}^\nu-
\frac{1}{4}{F_\mu}^\kappa{F_\kappa}^\lambda{F_\lambda}^\nu=0.
 \label{kinkDF=0}
\end{equation}
The maximal rank assumption implies the existence of an inverse
${{F^{-1}}_\mu}^\nu$ of the Kaluza-Klein gauge curvature, ${F_\mu}^\nu$,
${F_\mu}^\kappa{{F^{-1}}_\kappa}^\nu=
{{F^{-1}}_\mu}^\kappa{F_\kappa}^\nu=\delta_\mu^\nu$. Contracting
(\ref{kinkDF=0}) with ${{F^{-1}}_\nu}^\xi$ and rearranging terms we obtain
\begin{equation}
\frac{1}{4}{F_\mu}^\kappa{F_\kappa}^\xi= \left(k+
\frac{1}{8}F^2\right)\delta_\mu^\xi.
 \label{kinkDF=0'}
\end{equation}
Contraction eventually fixes the value of the constant to
$k=-\frac{d+2}{8d}F^2$. A covariantly constant gauge curvature
$F_{\mu\nu}$ is therefore solution of Grumiller-Jackiw equations
if and only if
\begin{equation}
{F_\mu}^\kappa{F_\kappa}^\nu= - \frac{1}{d}F^2\delta_\mu^\nu.
 \label{kinkDF=0''}
\end{equation}
Depending on the sign of $F^2$, $\mbox{sign}\{F^2\}\equiv\sigma$, these
equations introduce different kinds of spacetime structure, which are not
frequently encountered in theoretical physics, but are well studied in
differential geometry. By rescaling the Kaluza-Klein gauge curvature
${F_\mu}^\nu$, we introduce the mixed tensor
\begin{equation}
{J_\mu}^\nu=\pm\sqrt{\frac{d}{|F^2|}}{F_\mu}^\nu.
 \label{Almost(Para)ComplexStructure}
\end{equation}
Equation (\ref{kinkDF=0''}), the anti-symmetry of $F_{\mu\nu}$ and
(\ref{DF=0}) are then rewritten as
\begin{subequations}\label{(para)KaehlerStructure}
\begin{eqnarray}
&&{J_\mu}^\kappa{J_\kappa}^\nu= -\sigma \delta_\mu^\nu, \label{Almost(para)Complex}\\
&&{J_\mu}^\kappa{J_\nu}^\lambda g_{\kappa\lambda}=\sigma g_{\mu\nu},\label{(anti)HermitianMetric}\\
&&D_\kappa{J_\mu}^\nu=0\label{(para)KaehlerIntegrability}.
\end{eqnarray}
\end{subequations}
For $\sigma=+$ equation (\ref{Almost(para)Complex}) identifies ${J_\mu}^\nu$
with an \emph{almost complex structure} on spacetime,
(\ref{(anti)HermitianMetric}) states that $g_{\mu\nu}$ is an associated
\emph{Hermitian metric}, while (\ref{(para)KaehlerIntegrability}) guarantees
the integrability of the structure, making spacetime a
\emph{pseudo-K\"ahler manifold} \cite{Yano64,Barros&Romero82}. This implies
that the even dimensional spacetime carries an even index $2\textrm{s}$,
$\textrm{s}=0,...,\textrm{d}$. No solutions with Lorentzian signature are
admitted. For $\sigma=-$ equation (\ref{Almost(para)Complex}) identifies
${J_\mu}^\nu$ with an
\emph{almost product structure}|more precisely an \emph{almost
para-complex structure}|on spacetime, (\ref{(anti)HermitianMetric}) states
that $g_{\mu\nu}$ is an associated \emph{anti-Hermitian metric}, while
(\ref{(para)KaehlerIntegrability}) again guarantees the integrability of the
structure, making spacetime a~\emph{para-K\"ahler manifold}
\cite{Yano64,Cruceanu&Fortuny&Gadea96}. This implies that spacetime
carries a neutral signature $\textrm{d}=d/2$. Inserting (\ref{kinkDF=0''})
in (\ref{Riemann}) and reexpressing everything in terms of ${J_\mu}^\nu$ we
obtain
\begin{eqnarray}
&&R_{\mu\nu\kappa\lambda}=\frac{F^2}{4d}
\left(g_{\mu\lambda}g_{\kappa\nu}-g_{\mu\kappa}g_{\lambda\nu}
+\sigma J_{\mu\lambda}J_{\nu\kappa}-\sigma J_{\mu\kappa}J_{\nu\lambda}
-2\sigma
J_{\mu\nu}J_{\kappa\lambda}\right).\label{const.(para)hol.sec.curv.}
\end{eqnarray}
This reveals that spacetime is a pseudo-K\"ahler manifold with constant
holomorphic sectional curvature, when $\sigma=+$ (see Proposition 2.1.\ and
Corollary 2.2.\ in Ref.~\cite{Barros&Romero82}) or a para-K\"ahler manifold
with constant para-holomorphic sectional curvature, when $\sigma=-$ (see
Propositions 3.7.\ and Theorem 3.8.\ in
Ref.~\cite{Gadea&AmilibiaMontesinos89}). When complete, spacetime is then a
\emph{complex}/\emph{para-complex space form}, the
complex/para-complex analogue of real space forms.\\
 The simplest examples of such spaces are provided by the
pseudo-Euclidean complex algebra
$\mathbb{C}^\textrm{d}_\textrm{s}$ and para-complex algebra
$\mathbb{A}^\textrm{d}$ of vanishing holomorphic, respectively,
para-holomorphic sectional curvature.\footnote{The use is that of
displaying the complex dimension $\textrm{d}$ and signature
$\textrm{s}$ for complex spaces and the para-complex dimension
$\textrm{d}$, but not the para-complex signature|which always
equals half of the dimension|for para-complex spaces.} They are
constructed by endowing $\mathbb{R}^\textrm{2d}$ with the metric
and the almost complex/para-complex structure
\begin{equation}
\eta_{\mu\nu}=\left(
\begin{array}{cc}
\sigma
\mbox{\boldmath{$\eta$}}_\textrm{d}&0\\
0&\mbox{\boldmath{$\eta$}}_\textrm{d}
\end{array}\right),\hskip0,8cm
{\varepsilon_\mu}^\nu=\left(
\begin{array}{cc}
0&\mbox{\boldmath{$\eta$}}_\textrm{d}\\
-\sigma\mbox{\boldmath{$\eta$}}_\textrm{d}&0
\end{array} \right),\label{gj}
\end{equation}
with $\mbox{\boldmath{$\eta$}}_\textrm{d}$ the matrix corresponding to a
real $\textrm{d}$-dimensional pseudo-Euclidean metric carrying arbitrary
signature. On the other hand, it is readily checked that
$g_{\mu\nu}=\eta_{\mu\nu}$ and ${F_\mu}^\nu\propto{\varepsilon_\mu}^\nu$ are
solutions of equations (\ref{GJ}) only if $F^2=0$, so that
$\mathbb{C}^\textrm{d}_\textrm{s}$ and $\mathbb{A}^\textrm{d}$ can|in the
best case|only be enumerated among null rank solutions. For a non-vanishing
$F^2$ the theorems mentioned above identify the constant
holomorphic/para-holomorphic sectional curvature with $\frac{F^2}{d}$.
Indefinite complex space forms ($\sigma=+$) of non-vanishing holomorphic
sectional curvature were investigated by Barros and
Romero~\cite{Barros&Romero82}. They are locally isomorphic to the complex
pseudo-projective space $\mathbb{C}P^\textrm{d}_\textrm{s}$ with positive
holomorphic sectional curvature or to the complex pseudo-hyperbolic space
$\mathbb{C}H^\textrm{d}_\textrm{s}$ with negative holomorphic sectional
curvature---one is obtained by the other by replacing the metric with its
opposite. Para-complex space forms ($\sigma=-$) of non-vanishing
para-holomorphic sectional curvature were instead constructed by Gadea and
Montesinos~Amilibia~\cite{Gadea&AmilibiaMontesinos89} and further
investigated by Gadea and Mu\~{n}oz~Masqu\'{e}~\cite{Gadea&MunozMasque92}.
They are locally isomorphic to the para-complex projective model
$\mathbb{B}P^\textrm{d}$ with positive para-holomorphic sectional curvature
or to the para-complex hyperbolic model $\mathbb{B}H^\textrm{d}$ with
negative para-holomorphic sectional curvature|once again, one is obtained by
the other by changing the sign of the metric.\footnote{Gadea and
Montesinos~Amilibia introduce
\emph{para-complex projective models} $P_\textrm{d}(\mathbb{B})$
carrying both positive and negative para-holomorphic sectional curvature. We
partially modify their notation and distinguish projective
$\mathbb{B}P^\textrm{d}\equiv P_\textrm{d}(\mathbb{B})$|for positive
para-holomorphic sectional curvature|from hyperbolic
$\mathbb{B}H^\textrm{d}\equiv P_\textrm{d}(\mathbb{B})$|for negative
para-holomorphic sectional curvature|para-complex models to conform to real
and complex space forms.} The explicit form of the metric, $g_{\mu\nu}$, and
of the vector potential, $A_\mu$, generating $F_{\mu\nu}$ and hence
${J_\mu}^\nu$, are obtained as
\begin{subequations}
\begin{eqnarray}
&&g_{\mu\nu}=\frac{1}{\left(1+\frac{F^2}{4d}%|x|^2
\eta_{\kappa\lambda}x^\kappa x^\lambda
\right)^2}\left[\eta_{\mu\nu}+
\frac{F^2}{4d}\left(\eta_{\mu\nu}\eta_{\kappa\lambda}
-\eta_{\mu\kappa}\eta_{\nu\lambda}-\sigma
\varepsilon_{\mu\kappa}\varepsilon_{\nu\lambda} \right)x^\kappa
x^\lambda \right],
\nonumber\\
\label{g_const.(para)hol-sec.curv.}\\
&&A_\mu= \pm\sqrt{\frac{|F^2|}{d}} \left(1+\frac{F^2}{4d}%|x|^2
\eta_{\kappa\lambda}x^\kappa x^\lambda
\right){\varepsilon_\mu}^\nu g_{\nu\kappa} x^\kappa,
\label{A_const.(para)hol-sec.curv.}
\end{eqnarray}\label{g,A_const.(para)hol-sec.curv.}\end{subequations}
with  $\eta_{\mu\nu}$, ${\varepsilon_\mu}^\nu$ given by (\ref{gj}) and
$\varepsilon_{\mu\nu}={\varepsilon_\mu}^\kappa\eta_{\kappa\nu}$. We remark
that complex/para-complex space forms are neither spaces of constant
curvature nor conformally flat spaces. By direct substitution of
(\ref{g,A_const.(para)hol-sec.curv.}) in (\ref{GJ}) it is possible to check
that
\begin{equation}
\mathbb{C}P^\textrm{d}_\textrm{s}\left(\frac{|F^2|}{d}\right)\hskip0.5cm
\mbox{and}\hskip0.5cm
\mathbb{B}H^\textrm{d}\left(-\frac{|F^2|}{d}\right)
\end{equation}
|in brackets we give the holomorphic/para-holomorphic sectional curvature|
are indeed solutions of the Grumiller-Jackiw equations corresponding to a
positive signature for the extra coordinate $x^d$. Since the replacement of
the higher dimensional metric with its opposite produces a change in sign of
$g_{\mu\nu}$ and hence the replacement of projective with hyperbolic spaces
and viceversa, the remaining space forms
\begin{equation}
\mathbb{C}H^\textrm{d}_\textrm{s}\left(-\frac{|F^2|}{d}\right)\hskip0.5cm
\mbox{and}\hskip0.5cm
\mathbb{B}P^\textrm{d}\left(\frac{|F^2|}{d}\right)
\end{equation}
are instead solutions of the Grumiller-Jackiw equations
corresponding to  a negative signature for the extra coordinate
$x^d$.

 Real, complex and para-complex space forms are therefore seen
under the same light as solutions of the equations describing  the
Kaluza-Klein reduction of conformally flat spaces. It is then natural to
wonder what other spacetime structures fulfill Grumiller-Jackiw equations.
As far as maximal rank solutions are concerned, we observe that no extra
solutions can be constructed by conformal deformation of
pseudo-K\"ahler/para-K\"ahler structures. In fact, for $d>2$, the closure
condition $(dF)_{\mu\nu\kappa}=0$ immediately implies the constancy of the
conformal factor. Nor can extra solutions be obtained from almost
complex/para-complex structures by relaxing the integrability condition
(\ref{(para)KaehlerIntegrability}). In fact, the identity $D_\nu
F_{\mu\kappa}F^{\nu\kappa}= \frac{d}{4}D_\mu F^2$, obtained in
\S\ref{RRScurvatures}, is compatible with (\ref{kinkDF=0''}) if and only if
$F^2$ is constant or $d=2$. For these reasons we suspect the
complex/para-complex space forms to be the only maximal rank solutions of
equations (\ref{GJ}), but we could not prove this statement.

\section{Intermediate rank solutions}
\label{Int}

Next we consider the case in which the Kaluza-Klein gauge field $F_{\mu\nu}$
has intermediate rank $0<\mbox{rank}\{F_{\mu\nu}\}\equiv r<d$ and  nullity
$\mbox{null}\{F_{\mu\nu}\}=d-r\equiv n$. Given the closure condition,
$(dF)_{\mu\nu\kappa}=0$, a classical theorem of Darboux\footnote{Darboux
theorem further ensures the possibility of setting
$\textrm{F}_{\alpha\beta}$ in a canonical  form. This is, however, of no
relevance in our analysis.} ensures the possibility of finding, in a finite
neighborhood of every point, local coordinates
$x^\mu=\left(\xi^\alpha,y^i\right)$ with $\alpha=0,...,r-1$, $i=1,...,n$, in
such a way that
\begin{equation}
F_{\mu\nu}=\left(
\begin{array}{cc}
\textrm{F}_{\alpha\beta} & 0 \\
0 & 0
\end{array}\right),\label{red.field}
\end{equation}
with
$\textrm{F}_{\alpha\beta}=\partial_\alpha\textrm{A}_\beta-
\partial_\beta\textrm{A}_\alpha$
an $r$-dimensional non-degenerate closed two-form. The $\xi^\alpha$ and
$y^i$ parameterize non-degenerate and null gauge field directions and will
be referred  as {\em external} and {\em internal} coordinates, respectively.
Given the antisymmetry of $\textrm{F}_{\alpha\beta}$ the external dimension
is always an even number, $r=2\textrm{r}$. Adapted coordinates are defined
up to the coordinate transformations $\xi^\alpha\rightarrow
\xi'^\alpha(\xi)$, $y^i\rightarrow y'^i(\xi,y)$, with internal
diffeomorphisms allowed to depend on external coordinates. In such
adapted frames the spacetime metric can be parameterized without
loss of generality as
\begin{equation}
g_{\mu\nu}=\left(
\begin{array}{cc}
\textrm{g}_{\alpha\beta}+  a_\alpha^k a_\beta^l h_{kl}
& a_\alpha^k h_{kj}\\
h_{il} a_\beta^l & h_{ij}
\end{array}\right), \label{red.metric}
\end{equation}
with $\textrm{g}_{\alpha\beta}$, $h_{ij}$ and $a_\alpha^i$ depending, in
general, on external and internal coordinates. Under the transformations
above $\textrm{g}_{\alpha\beta}$ and $h_{ij}$ transform as external and
internal metric tensors, respectively, while $a_\alpha^i$ identifies with an
external gauge potential taking values in the internal diffeomorphisms
algebra. The coordinate splitting is completely characterized by the lower
dimensional tensors
\begin{equation}
\hat{E}_{i\alpha\beta}=\frac{1}{2}\left(\partial_i
\textrm{g}_{\alpha\beta}+ f_{i\alpha\beta}\right),\hskip0,3cm
E_{\alpha ij} = \frac{1}{2} \left(\partial_\alpha h_{ij} -{\cal
L}_{a_\alpha}h_{ij}\right),
 \label{FundamentalForms}
\end{equation}
with $f^i_{\alpha\beta}=
\partial_\alpha a_\beta^i -
\partial_\beta a_\alpha^i -a_\alpha^j\partial_j a_\beta^i +
a_\beta^j\partial_j a_\alpha^i$ the gauge curvature associated to
the external vector potential $(a^i)_\alpha$ and ${\cal
L}_{a_\alpha}$ the Lie derivative with respect to the internal
vector $(a_\alpha)^i$. $\hat{E}_{i\alpha\beta}$ is a
\emph{generalized second fundamental form} for the external
space, which is not in general a spacetime submanifold. Most remarkable, the
vanishing of its antisymmetric part, $f^i_{\alpha\beta}$, ensures the
possibility of introducing internal coordinates in such a way that
$a_\alpha^i$ and, hence, the off-diagonal components of the $d$-dimensional
metric vanish identically. For every fixed value $\bar{\xi}$ of the external
coordinates, $E_{\alpha ij}|_{\xi=\bar{\xi}}$ represents instead the
standard second fundamental form of the corresponding internal space, which
is always a spacetime submanifold~\cite{Maraner&Pachos08}.

In the adapted coordinate frame equations (\ref{GJ3}) can be rewritten in
terms of the residual gauge field and the generalized fundamental forms as
\begin{subequations}\label{reducedGJ3}\begin{eqnarray}
&& \hat\nabla_\gamma \textrm{F}_{\alpha\beta}+\frac{2}{d-1}
\left(\textrm{g}_{\gamma[\alpha}\hat\nabla_\delta
{\textrm{F}_{\beta]}}^\delta+\textrm{g}_{\gamma[\alpha}
{\textrm{F}_{\beta]}}^\delta {E_{\delta a}}^a\right)=0,\label{RGJ3a}\\
&& {\textrm{F}_\alpha}^\delta\hat E_{j\gamma\delta}-
\frac{1}{d-1}\textrm{g}_{\alpha\gamma}\textrm{F}^{\delta\epsilon}
\hat E_{j\delta\epsilon}=0,\label{RGJ3b}\\
&& \nabla_k \textrm{F}_{\alpha\beta}=0,\label{RGJ3c}\\
&& {\textrm{F}_{\alpha}}^\delta E_{\delta jk}-\frac{1}{d-1}
\left(\hat\nabla_\delta {\textrm{F}_{\alpha}}^\delta+
{\textrm{F}_{\alpha}}^\delta {E_{\delta a}}^a\right)h_{jk}=0,\label{RGJ3d}\\
&&h_{k[i}\textrm{F}^{\gamma\delta}\hat{E}_{j]\gamma\delta}=0,\label{RGJ3e}
\end{eqnarray}\end{subequations}
with the relevant definition of the hatted derivative
$\hat\nabla_\alpha$ given below and $\nabla_i$ the standard
internal covariant derivative associated to $h_{ij}$.\footnote{The
general definition of $\hat\nabla_\alpha$ (Eq.(35) in
Ref.~\cite{Maraner&Pachos08}) is of no relevance here.} Equations
(\ref{RGJ3b}) and (\ref{RGJ3c}) immediately imply that $\partial_k
\textrm{g}_{\alpha\beta}=\hat{E}_{k\alpha\beta}+\hat{E}_{k\beta\alpha}=0$
and $\partial_k\textrm{F}_{\alpha\beta}=\nabla_k
\textrm{F}_{\alpha\beta}=0$, showing that the external metric and
the residual gauge field only depend on external coordinates
\begin{equation}
\textrm{g}_{\alpha\beta}=\textrm{g}_{\alpha\beta}(\xi),\hskip0.5cm
\textrm{F}_{\alpha\beta}=\textrm{F}_{\alpha\beta}(\xi).\label{gammaPhi}
\end{equation}
As a consequence, $\hat\nabla_\alpha$ coincides with the standard external
covariant derivative  associated to $\textrm{g}_{\alpha\beta}$,
$\hat\nabla_\alpha\equiv\nabla_\alpha$. Contracting (\ref{RGJ3a}) with
$\textrm{g}^{\beta\gamma}$, or (\ref{RGJ3d}) with $h^{jk}$, we obtain
$(r-1){\textrm{F}_{\alpha}}^\delta {E_{\delta i}}^i=n\nabla_\delta
{\textrm{F}_{\alpha}}^\delta$ or, equivalently,
\begin{equation}
{E_{\alpha i}}^i=\frac{n}{r-1} {{\textrm{F}^{-1}}_\alpha}^\beta
\nabla_\gamma {\textrm{F}_{\beta}}^\gamma,\label{traceFundForm}
\end{equation}
with ${{\textrm{F}^{-1}}_\alpha}^\beta$ the inverse of the
residual gauge curvature ${\textrm{F}_\alpha}^\beta$,
${\textrm{F}_\alpha}^\gamma{{\textrm{F}^{-1}}_\gamma}^\beta=
{{\textrm{F}^{-1}}_\alpha}^\gamma{\textrm{F}_\gamma}^\beta=\delta_\alpha^\beta$.
Substituting (\ref{traceFundForm}) back in (\ref{RGJ3a}) yields
\begin{equation}
\nabla_\gamma \textrm{F}_{\alpha\beta}+\frac{2}{r-1}
\textrm{g}_{\gamma[\alpha}\nabla_\delta
{\textrm{F}_{\beta]}}^\delta=0. \label{RGJ3a'}
\end{equation}
Equations (\ref{RGJ3a'}) precisely reproduce (\ref{GJ3}) on the external
subspace, i.e. along the non-degenerate directions of the Kaluza-Klein gauge
field. Substituting (\ref{traceFundForm}) back in (\ref{RGJ3d}) produces
instead
\begin{equation}
E_{\gamma ij}-\frac{1}{n}{E_{\gamma k}}^k h_{ij}=0,
 \label{RGJ3d'}
\end{equation}
implying that all internal spaces are totally umbilical and that
$(f_{\alpha\beta})^i$ is an internal conformal Killing vector, $\nabla_i
f_{j\alpha\beta}+\nabla_j f_{i\alpha\beta}=\frac{2}{n}(\nabla_k
f^k_{\alpha\beta})h_{ij}$ (see \S4.2. in Ref.~\cite{Maraner&Pachos08}). As a
consequence, it is always possible to further adapt internal coordinates in
such a way that the internal metric and the external gauge curvature
decompose as
\begin{equation}
h_{ij}=\lambda(\xi)\,c_{ij}(y), \hskip0.3cm
f^i_{\alpha\beta}=\textrm{f}_{\alpha\beta}^{\,\textsf{a}}(\xi)\,C^i_\textsf{a}(y),
\label{totalyumbilic}
\end{equation}
with $C^i_\textsf{a}$, $\textsf{a}=1,...,(n+1)(n+2)/2$, a basis of the
internal conformal algebra. Contracting now (\ref{RGJ3b}) with
$\textrm{g}^{\alpha\gamma}$, or (\ref{RGJ3e}) with $h^{ik}$, we eventually
obtain
\begin{equation}
(n-1)\textrm{F}^{\alpha\beta}\hat{E}_{i\alpha\beta}=0.
\label{RGJ3beC}
\end{equation}
For $n>1$ (\ref{RGJ3beC}) requires $\textrm{F}^{\alpha\beta}
\hat{E}^i_{\alpha\beta}=\frac{1}{2}\textrm{F}^{\alpha\beta}f^i_{\alpha\beta}=0$,
which substituted back in (\ref{RGJ3b}) implies the vanishing of
$f^i_{\alpha\beta}$. For $n=1$ (\ref{RGJ3beC}) is identically
satisfied and (\ref{RGJ3b}) reduces to a traceless equation
implying the proportionality between $f_{\alpha\beta} \equiv
f^1_{\alpha\beta}$ and the inverse Kaluza-Klein field
${\textrm{F}^{-1}}_{\alpha\beta}$; correspondingly the sum in
(\ref{totalyumbilic}) reduces to a single element of the internal
conformal algebra. Summarizing,
\begin{equation}
f^i_{\alpha\beta}=0 \hskip0.4cm\mbox{for}\hskip0.2cm n>1\hskip0.5cm
\mbox{and}\hskip0.5cm f_{\alpha\beta}=-\frac{1}{r} f_{\gamma\delta}
\textrm{F}^{\gamma\delta} {\textrm{F}^{-1}}_{\alpha\beta} \hskip0.4cm
\mbox{for}\hskip0.2cm n=1.\label{field-f}
\end{equation}
The cases $n>1$ and $n=1$ are, therefore, better treated separately.

\subsection{Nullity greater than one}
\label{n>1}

By means of the generalizations of Gauss, Codazzi and Ricci equations
\cite{Maraner&Pachos08}, that express higher dimensional curvatures in terms
of lower dimensional curvatures and generalized fundamental forms, it is
immediately possible to reduce (\ref{Riemann}) in its lower dimensional
components. Of the six resulting equations only two are not identically
satisfied
\begin{eqnarray}
&&\textrm{R}_{\alpha\beta\gamma\delta}=2\left(k+\frac{1}{8}\textrm{F}^2\right)
\gamma_{\alpha[\delta}\gamma_{\gamma]\beta}\nonumber\\
&&\hskip1.5cm-\frac{1}{2}\left(\gamma_{\alpha[\gamma}
\textrm{F}_{\delta]\xi}{\textrm{F}_\beta}^\xi-\gamma_{\beta[\gamma}\textrm{F}_{\delta]\xi}
{\textrm{F}_\alpha}^\xi\right)
-\frac{1}{2}\left(\textrm{F}_{\alpha\beta}\textrm{F}_{\gamma\delta}-
\textrm{F}_{\alpha[\gamma}\textrm{F}_{\delta]\beta}\right),\label{RiemannABCD n>1}\\
&&K_{ijkl}=2\left(k+\frac{1}{8}\textrm{F}^2+\frac{1}{n^2}{{E_\alpha}_m}^m{{E^\alpha}_n}^n\right)
h_{i[l}h_{k]j}, \label{Riemannabcd n>1}%\\
\end{eqnarray}
with $\textrm{R}_{\alpha\beta\gamma\delta}$ and $K_{ijkl}$ the Riemann
tensors associated to the external metric $\textrm{g}_{\alpha\beta}$ and the
internal metric $h_{ij}$, respectively, and
$\textrm{F}^2=\textrm{F}_{\alpha\beta}\textrm{F}^{\alpha\beta}=F^2$. The
Killing vector (\ref{Killing}) reduces to
$K_\mu=\left(\frac{1}{r-1}\nabla_\beta {\textrm{F}_{\alpha}}^\beta,
0\right)$. Correspondingly, the integrability conditions (\ref{kinkEq})
split in four lower dimensional equations. The only one which is
non-identically satisfied reads
\begin{eqnarray}
&&\frac{1}{2}\nabla^2 {\textrm{F}_\alpha}^\beta+\left(k+
\frac{1}{8}\textrm{F}^2\right){\textrm{F}_\alpha}^\beta-
\frac{1}{4}{\textrm{F}_\alpha}^\gamma{\textrm{F}_\gamma}^\delta{\textrm{F}_\delta}^\beta=0.
\label{kinkEq n>1}
\end{eqnarray}
We recognize that (\ref{RGJ3a'}), (\ref{RiemannABCD n>1}) and
(\ref{kinkEq n>1}) respectively reproduce (\ref{GJ3}),
(\ref{Riemann}) and (\ref{kinkEq}) when $r\rightarrow d$,
$\xi^\alpha\rightarrow x^\mu$ and
$\textrm{F}_{\alpha\beta}\rightarrow F_{\mu\nu}$. For $n>1$ the
problem along external directions is therefore fully equivalent to
finding maximal rank solutions of our original set of equations.
The only difference is that the residual rank $r$ is also allowed
to take the value $r=2$, precluded to the spacetime dimension $d$.
Once the external space geometry is determined by (\ref{RGJ3a'}),
(\ref{RiemannABCD n>1}), (\ref{kinkEq n>1}), equations
(\ref{Riemannabcd n>1}) fix the geometry of internal spaces
correspondingly.

\subsubsection{Equations (\ref{RGJ3a'}), (\ref{RiemannABCD n>1}), (\ref{kinkEq n>1}) for $r=2$}
 \label{r=2}
In two dimensions the gauge curvature is always proportional to the
invariant volume element, so that we can set in full generality
${\textrm{F}_\alpha}^\beta=\varphi\,{\varepsilon_\alpha}^\beta$, with
${\varepsilon_\alpha}^\beta$ given by (\ref{gj}). Equations
(\ref{RiemannABCD n>1}), (\ref{kinkEq n>1}) and the dimensional reduced
(\ref{traceless}) take then the form
\begin{subequations}\begin{eqnarray}
&&R=2k+3\sigma\varphi^2,\\
&&\nabla^2\varphi+2k\varphi+\sigma\varphi^3=0,\\
&&\nabla_\alpha\nabla_\beta\varphi-\frac{1}{2}\gamma_{\alpha\beta}\nabla^2\varphi=0,
\end{eqnarray}\label{GravitationalKinkEq}\end{subequations}
respectively, reproducing the `curvature constraint', `gravitational-kink'
and `traceless' equations obtained by Guralnik, Iorio, Jackiw and Pi from
the Kaluza-Klein reduction of the gravitational Chern-Simons term
((4.47,48,49) in Ref.~\cite{Guralnik&Iorio&Jackiw&Pi03}). Local solutions
are constructed in their paper and extended globally in
Ref.~\cite{Grumiller&Kummer03}. Besides the symmetry preserving solutions
$\mathbb{R}^2_s$, $S^2_s$, $H^2_s$ and the symmetry breaking solutions
$\mathbb{C}P^\textrm{1}_\textrm{s}$, $\mathbb{B}H^\textrm{1}$|and
$\mathbb{C}H^\textrm{1}_\textrm{s}$, $\mathbb{B}P^\textrm{1}$ for a negative
signature of $x^d$|for $\sigma=-$ and $k>0$ they found the extra class of
`gravitational kink' solutions
\begin{eqnarray}
&&\textrm{g}_{\alpha\beta}=\left(
\begin{array}{cc}
-k^2\,\mbox{sech}^4\left(\sqrt{\frac{k}{2}}\xi^1\right) &0\\
0&1
\end{array}
\right),\hskip0,5cm \textrm{A}_\alpha=\mbox{$\left(\pm
k\,\mbox{sech}^2\left(\sqrt{\frac{k}{2}}\xi^1\right),0\right)$},\label{Kink}
\end{eqnarray}
with the corresponding kink profile
\begin{equation}
\mbox{$\varphi(\xi)=
\pm\sqrt{2k}\tanh\left(\sqrt{\frac{k}{2}}\xi^1\right).$}
\label{KinkProfile}
\end{equation}
These solutions are associated to  para-K\"ahler structures defined on
spacetime. It is in fact easy to check that
${J_\alpha}^\beta=\pm{\textrm{F}_\alpha}^\beta/\varphi$, fulfills conditions
(\ref{(para)KaehlerStructure}) with $\sigma=-$. The scalar
$\textrm{F}^2=-2\varphi^2$ is however non-constant and the Killing vector is
correspondingly non-vanishing, $\nabla_\beta
{\textrm{F}_{\alpha}}^\beta=\left(k^2\
\textrm{sech}^4\sqrt{\frac{k}{2}}\xi^1, 0\right)$. The solutions
corresponding to a negative signature of the extra Kaluza-Klein coordinate
$x^d$ are obtained by replacing the metric with its opposite. In the latter
(former) case, for small values of $|\sqrt{k}\xi^1|$ the curvature is
negative (positive). For larger values it is positive (negative), achieving
$\textrm{dS}_2$ ($\textrm{AdS}_2$) at infinity. While the metric reproduces
asymptotically the deSitter (anti-deSitter) spacetime, the modulo of the
gauge field correspond to a kink profile. For these reasons it is natural to
refer to these spaces as \emph{kink} and
\emph{anti-kink} spaces. We denote them by $K^2_1(k)$ and
$\textrm{A}K^2_1(k)$ respectively, where we give in brackets the positive
parameter labelling the solution.

\subsubsection{$r\geq2$, $n>1$: solutions from complex/para-complex space forms}
 \label{n>1,r>=2}
The complex/para-complex space forms $\mathbb{C}P^\textrm{r}_\textrm{s}$,
$\mathbb{B}H^\textrm{r}$|and $\mathbb{C}H^\textrm{r}_\textrm{s}$,
$\mathbb{B}P^\textrm{r}$ for a negative signature of $x^d$|generate the
following intermediate rank solutions of Grumiller-Jackiw equations. For
every even, strictly positive value of $r=2\textrm{r}$ the external space is
a complex/para-complex space form of real dimension $r$ and constant
holomorphic/para-holomorphic sectional curvature $\frac{\textrm{F}^2}{r}$.
The Killing vector $K_\mu$ and the fundamental forms $E_{\alpha ij}$ vanish
identically, $\textrm{F}^2$ is constant and $k=-\frac{r+2}{8r}\textrm{F}^2$.
Consequently, equation (\ref{Riemannabcd n>1}) require the internal spaces
to be $n$-dimensional real space forms of sectional curvature
$-\frac{\textrm{F}^2}{4r}$. Spacetime results into the direct product of a
complex/para-complex space form of holomorphic/para-holomorphic sectional
curvature $\frac{\textrm{F}^2}{r}$ and a real space form of sectional
curvature $-\frac{\textrm{F}^2}{4r}$. For $\textrm{F}^2>0$ and
$\textrm{F}^2<0$ we respectively obtain
\begin{equation}
\mathbb{C}P^\textrm{r}_\textrm{s}\left(\frac{|\textrm{F}^2|}{r}\right)
\times\mathbb{R}H^n_{s}\left(-\frac{|\textrm{F}^2|}{4r}\right)\hskip0,3cm
\mbox{and}\hskip0,3cm
\mathbb{B}H^\textrm{r}\left(-\frac{|\textrm{F}^2|}{r}\right)
\times \mathbb{R}P^n_{s}\left(\frac{|\textrm{F}^2|}{4r}\right),
\end{equation}
with external and internal signatures unrelated. The choice of a
negative signature for the extra coordinate $x^d$ produce instead
the solutions
\begin{equation}
\mathbb{C}H^\textrm{r}_\textrm{s}\left(-\frac{|\textrm{F}^2|}{r}\right)
\times
\mathbb{R}P^n_{s}\left(\frac{|\textrm{F}^2|}{4r}\right)\hskip0,3cm
\mbox{and}\hskip0,3cm
\mathbb{B}P^\textrm{r}\left(\frac{|\textrm{F}^2|}{r}\right) \times
\mathbb{R}H^n_{s}\left(-\frac{|\textrm{F}^2|}{4r}\right).
\end{equation}
Metrics and vector potentials are immediately constructed by means
of (\ref{g,A_const.(para)hol-sec.curv.}) and
(\ref{g,A_const.sec.curv.}).

\subsubsection{$r=2$, $n>1$: kinks}
 \label{n>1,r=2}
The exceptional class of rank two kink/anti-kink solutions
discussed in \S\ref{r=2} also generates intermediate rank
solutions of Grumiller-Jackiw equations. For a positive signature
of the Kaluza-Klein extra coordinate $x^d$, the external space
metric is that of an anti-kink space $\textrm{A}K^2_1(k)$. Since
in two dimensions the squared gauge field is always proportional
to delta, ${\textrm{F}_\alpha}^\gamma {\textrm{F}_\gamma}^\beta=-
\frac{1}{2} \textrm{F}^2\delta_\alpha^\beta$, from
(\ref{traceFundForm}) and the fundamental form definition
(\ref{FundamentalForms}) it is possible to show that the scale factor
$\lambda(\xi)$ appearing in the internal metric is always proportional to
the squared field modulo $\textrm{F}^2$. Up to a multiplicative constant we
therefore have
\begin{equation}
\lambda(\xi)=\pm\;4k\;\mbox{$\tanh^2\sqrt{\frac{k}{2}}\xi^1$}.
\label{warpfactor}
\end{equation}
Equation (\ref{Riemannabcd n>1}) fixes then the internal spaces to be
$n$-dimensional real space forms of constant sectional curvature $\pm2k^2$.
For a positive choice of the warp factor spacetime results into the warped
product of the anti-kink space $\textrm{A}K^2_1(k)$ and the pseudo-sphere
$\mathbb{R}P^n_s\left(2k^2\right)$
\begin{equation}
 \textrm{A}K^2_1(k)\times_{4k\tanh^2\sqrt{\frac{k}{2}}\xi^1}\mathbb{R}P^n_s\left(2k^2\right),
\end{equation}
while for a negative choice the second term is replaced by the
pseudo-hyperbolic space $\mathbb{R}H^n_s\left(2k^2\right)$
\begin{equation}
 \textrm{A}K^2_1(k)\times_{-4k\tanh^2\sqrt{\frac{k}{2}}\xi^1}\mathbb{R}H^n_s\left(-2k^2\right).
\end{equation}
The solutions corresponding to a negative signature of the extra
Kaluza-Klein coordinate are obtained by changing the sign of the
higher dimensional metric. For every positive value of $k$
spacetime results into the warped product of the kink space
$K^2_1(k)$ with either the pseudo-hyperbolic space
$\mathbb{R}H^n_s\left(-2k^2\right)$
\begin{equation}
\textrm{K}^2_1(k)\times_{4k\tanh^2\sqrt{\frac{k}{2}}\xi^1}\mathbb{R}H^n_s\left(-2k^2\right),
\end{equation}
or the pseudo-sphere $\mathbb{R}P^n_s\left(2k^2\right)$
\begin{equation}
\textrm{K}^2_1(k)\times_{-4k\tanh^2\sqrt{\frac{k}{2}}\xi^1}\mathbb{R}P^n_s\left(2k^2\right).
\end{equation}
Explicit expressions of metrics and vector potentials are
immediately constructed by means of (\ref{Kink}),
(\ref{g,A_const.sec.curv.}) and (\ref{totalyumbilic}).

\subsection{Nullity equal to one}
\label{n=1}

Eventually, we consider solutions with $r=d-1$ and $n=1$. This is
the only case in which it is not in general  possible to introduce
coordinates bringing the spacetime metric (\ref{red.metric}) in
block-diagonal form. In different words, this is the only case in
which the gauge field $f_{\alpha\beta}$ can be different than
zero. It is convenient to rescale the internal coordinate in such
a way that $h_{11}=\lambda(\xi)$ and set $\lambda
f_{\alpha\beta}\textrm{F}^{\alpha\beta}=2rl$, with $l(\xi,y)$ some
undetermined function of the coordinates. The Riemann tensor
(\ref{Riemann}) is then again reduced by means of generalized
Gauss, Codazzi and Ricci equations. Of the resulting conditions
only one is not identically satisfied. Taking (\ref{field-f}) into
account it reads
\begin{eqnarray}
&&\textrm{R}_{\alpha\beta\gamma\delta} =
2\left(k+\frac{1}{8}\textrm{F}^2\right)
\gamma_{\alpha[\delta}\gamma_{\gamma]\beta}
-\frac{1}{2}\left(\gamma_{\alpha[\gamma}
\textrm{F}_{\delta]\xi}{\textrm{F}_\beta}^\xi-\gamma_{\beta[\gamma}\textrm{F}_{\delta]\xi}
{\textrm{F}_\alpha}^\xi\right)\nonumber\\&&\hskip0.7cm
-\frac{2l^2}{\lambda}
\left({\textrm{F}^{-1}}_{\alpha\beta}{\textrm{F}^{-1}}_{\gamma\delta}-
{\textrm{F}^{-1}}_{\alpha[\gamma}{\textrm{F}^{-1}}_{\delta]\beta}\right)
-\frac{1}{2}\left(\textrm{F}_{\alpha\beta}\textrm{F}_{\gamma\delta}-
\textrm{F}_{\alpha[\gamma}\textrm{F}_{\delta]\beta}\right),
 \label{RiemannABCD n=1}
\end{eqnarray}
with $\textrm{R}_{\alpha\beta\gamma\delta}$ again denoting the
Riemann tensor associated to $\textrm{g}_{\alpha\beta}$. The
Killing vector (\ref{Killing}) reduces now to
$K_\mu=\left(\frac{1}{r-1} \nabla_\beta
{\textrm{F}_{\alpha}}^\beta+ l\,a^1_\alpha, l \right)$. Eventually, the integrability condition
(\ref{kinkEq}) yields the lower dimensional equations
\begin{eqnarray}
&&\frac{1}{2}\nabla^2 {\textrm{F}_\alpha}^\beta
 +\frac{l^2}{\lambda}{{\textrm{F}^{-1}}_\alpha}^\beta+\left(k+
\frac{1}{8}\textrm{F}^2\right){\textrm{F}_\alpha}^\beta-
\frac{1}{4}{\textrm{F}_\alpha}^\gamma{\textrm{F}_\gamma}^\delta{\textrm{F}_\delta}^\beta=0,
\label{kinkEq n=1a}\\
&& \nabla_\alpha\, l=0 \hskip0,3cm\mbox{and}\hskip0,3cm \nabla_y\,
l=0. \label{kinkEq n=1b}
\end{eqnarray}
The first is the integrability condition for (\ref{RGJ3a'}) subject to
(\ref{RiemannABCD n=1}). The other two fix $l$ to a constant. For $l=0$
equations (\ref{RiemannABCD n=1}), (\ref{kinkEq n=1a}) exactly reproduce
(\ref{RiemannABCD n>1}), (\ref{kinkEq n>1}), or equivalently,
(\ref{Riemann}), (\ref{kinkEq}). As a consequence every maximal rank
solution, including rank two, generates a nullity one solution. Proceeding
as in
\S\ref{n>1,r>=2} and \S\ref{n>1,r=2}, for a positive signature of
the extra Kaluza-Klein coordinate, we obtain the  solutions
\begin{equation}
\mathbb{C}P^\textrm{r}_\textrm{s}\left(\frac{|\textrm{F}^2|}{r}\right)\times\mathbb{R},\hskip0.3cm
\mathbb{B}H^\textrm{r}\left(-\frac{|\textrm{F}^2|}{r}\right)\times\mathbb{R},
\end{equation}
together with the anti-kink warped products
\begin{equation}
\textrm{A}K^2_1(k)\times_{\pm4k\tanh^2\sqrt{\frac{k}{2}}\xi^1}\mathbb{R}.\label{AKinkR}
\end{equation}
For a negative choice of the extra coordinate we have instead
\begin{equation}
\mathbb{C}H^\textrm{r}_\textrm{s}\left(-\frac{|\textrm{F}^2|}{r}\right)\times\mathbb{R},\hskip0.3cm
\mathbb{B}P^\textrm{r}\left(\frac{|\textrm{F}^2|}{r}\right)\times\mathbb{R},
\end{equation}
with the kink warped products
\begin{equation}
K^2_1(k)\times_{\pm4k\tanh^2\sqrt{\frac{k}{2}}\xi^1}\mathbb{R}.\label{KinkR}
\end{equation}
For $l\neq0$ new terms appear in the Riemannian curvature
(\ref{RiemannABCD n=1}) and in the integrability condition
(\ref{kinkEq n=1a}) and some extra consideration is necessary.

\subsubsection{$r\geq2$, $n=1$: more solutions from complex/para-complex space forms}
 \label{n=1,r>=2}
Given the structure of the extra terms in (\ref{RiemannABCD n=1})
and (\ref{kinkEq n=1a}), it is natural to look for solutions
related to K\"ahler and para-K\"ahler structures by a constant
rescaling
\begin{equation}
{\textrm{F}_\alpha}^\beta=\pm\sqrt{\frac{|\textrm{F}^2|}{r}}{J_\alpha}^\beta,
\label{Ansatz}
\end{equation}
where ${J_\alpha}^\beta$ fulfills conditions
(\ref{(para)KaehlerStructure}) with $\sigma$ the sign of
$\textrm{F}^2$ and where the constant of proportionality has been
fixed by squaring and tracing both members of the equality. The
constancy of $\textrm{F}^2$ implies the constancy of
$\lambda(\xi)$ which is set to plus or minus one by a proper
rescaling of the internal coordinate, $\lambda=\pm1$. Equations
(\ref{Ansatz}) and (\ref{(para)KaehlerStructure}) fix the value of
the inverse Kaluza-Klein curvature to
\begin{equation}
{{\textrm{F}^{-1}}_\alpha}^\beta=-\frac{r}{\textrm{F}^2}\,{\textrm{F}_\alpha}^\beta.
 \label{F^-1}
\end{equation}
By substituting (\ref{F^-1}) back in (\ref{kinkEq n=1a}),
recalling that (\ref{(para)KaehlerIntegrability}) requires the
vanishing of $\nabla^2 {\textrm{F}_\alpha}^\beta$ and proceeding
as in \S\ref{NullMax}, the integrability conditions fix the value
of the constant to
$k=\frac{rl^2}{\lambda\textrm{F}^2}-\frac{(r+2)\textrm{F}^2}{8r}$.
The eventual substitution of (\ref{F^-1}) and $k$ in
(\ref{RiemannABCD n=1}) yields the Riemann tensor
\begin{eqnarray}
&&\textrm{R}_{\alpha\beta\gamma\delta}=
\left(\frac{\textrm{F}^2}{2r}+\frac{2rl^2}{\lambda\textrm{F}^2}\right)
 \left(
  \textrm{g}_{\alpha[\delta}\textrm{g}_{\gamma]\beta}
 -\sigma J_{\alpha[\delta}J_{\gamma]\beta}
 -\sigma J_{\alpha\beta}J_{\gamma\delta}
  \right),\label{const.(para)hol.sec.curv. n=1}
\end{eqnarray}
showing that the external space is either a pseudo-K\"ahler or a
para-K\"ahler manifold with constant holomorphic, respectively,
para-holomorphic sectional curvature
$\frac{\textrm{F}^2}{r}+\frac{4rl^2}{\lambda\textrm{F}^2}$. Taking
(\ref{red.metric}) and (\ref{field-f}) into account, we see that spacetime
results itself in a Kaluza-Klein space, with external space given by a
complex/para-complex space form and gauge structure proportional to the
underlying complex/paracomplex structure
\begin{equation}
f_{\alpha\beta}=\pm2l\,\sqrt{\frac{r}{|\textrm{F}^2|}}\,J_{\alpha\beta}.
\label{gf n=1}
\end{equation}
When $\frac{\textrm{F}^2}{r}+\frac{4rl^2}{\lambda\textrm{F}^2}=0$ the
underlying space form has zero holomrphic/para-holomorphic sectional
curvature, corresponding to $\mathbb{C}^\textrm{r}_\textrm{s}$ for
$\textrm{F}^2>0$ and to $\mathbb{A}^\textrm{r}$ for $\textrm{F}^2<0$.
Missing a standard notation, we borrow and slightly modify the warped
product notation and denote these `Kaluza-Klein products' as
\begin{equation}
\mathbb{C}^\textrm{r}_\textrm{s}\left(0\right)\times^{\pm
\sqrt{\frac{|\textrm{F}^2|}{r}}J_{\alpha\beta}}\mathbb{R}
 \hskip0.3cm\mbox{and}\hskip0.3cm
\mathbb{A}^\textrm{r}\left(0\right)\times^{\pm
\sqrt{\frac{|\textrm{F}^2|}{r}}J_{\alpha\beta}}\mathbb{R}.
\label{KKsc=0}
\end{equation}
When $\frac{\textrm{F}^2}{r}+\frac{4rl^2}{\lambda\textrm{F}^2}>0$ the space
form has positive holomorphic/para-holomorphic sectional curvature,
corresponding to $\mathbb{C}P^\textrm{r}_\textrm{s}$ for $\textrm{F}^2>0$
and to $\mathbb{A}^\textrm{r}P$ for $\textrm{F}^2<0$. The corresponding
spaces are
\begin{equation}
\mathbb{C}P^\textrm{r}_\textrm{s}\left(
\mbox{$\frac{\textrm{F}^2}{r}+\frac{4rl^2}{\lambda\textrm{F}^2}$}
\right)\times^{\pm2l\sqrt{\frac{r}{|\textrm{F}^2|}}J_{\alpha\beta}}\mathbb{R}
 \hskip0.3cm\mbox{and}\hskip0.3cm
\mathbb{B}P^\textrm{r}\left(
\mbox{$\frac{\textrm{F}^2}{r}+\frac{4rl^2}{\lambda\textrm{F}^2}$}
\right)\times^{\pm2l\sqrt{\frac{r}{|\textrm{F}^2|}}J_{\alpha\beta}}\mathbb{R}.
\label{KKsc>0}
\end{equation}
Eventually,  when
$\frac{\textrm{F}^2}{r}+\frac{4rl^2}{\lambda\textrm{F}^2}<0$ the
holomrphic/para-holomorphic sectional curvature is negative and
the space form  corresponds to $\mathbb{C}H^\textrm{r}_\textrm{s}$
for $\textrm{F}^2>0$ and to $\mathbb{A}^\textrm{r}H$ for
$\textrm{F}^2<0$. The relative solutions are
\begin{equation}
\mathbb{C}H^\textrm{r}_\textrm{s}\left(
\mbox{$\frac{\textrm{F}^2}{r}+\frac{4rl^2}{\lambda\textrm{F}^2}$}
\right)\times^{\pm2l\sqrt{\frac{r}{|\textrm{F}^2|}}J_{\alpha\beta}}\mathbb{R}
 \hskip0.3cm\mbox{and}\hskip0.3cm
\mathbb{B}H^\textrm{r}\left(
\mbox{$\frac{\textrm{F}^2}{r}+\frac{4rl^2}{\lambda\textrm{F}^2}$}
\right)\times^{\pm2l\sqrt{\frac{r}{|\textrm{F}^2|}}J_{\alpha\beta}}\mathbb{R}.
\label{KKsc<0}
\end{equation}
Explicit forms of metrics and gauge fields are immediately obtained by means
of (\ref{g,A_const.(para)hol-sec.curv.}), (\ref{red.metric}) and (\ref{gf
n=1}). The choice of a negative signature for the extra Kaluza-Klein
coordinate produces exactly the same solutions.

\subsubsection{$r=2$, $n=1$: kinks centripetal/centrifugal deformations}
 \label{n=1,r=2}
\begin{figure}[t]
 \vspace*{8pt}
 \centerline{
\psfig{file=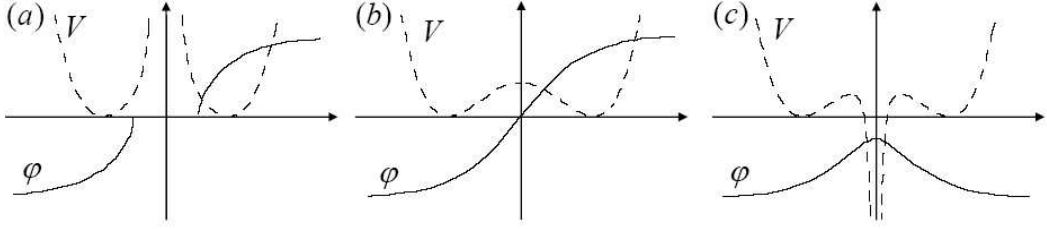,width=14.0cm}\newline
 }
 \vspace*{8pt}
 \caption{The gravitational kink profile with centripetal deformation (solid line) (a),
without deformation (b) and with centripetal deformation (c), together with
the corresponding potentials (dashed line).}
 \label{figure}
\end{figure}

For two external dimensions we can set again in full generality
${\textrm{F}_\alpha}^\beta=\varphi\,{\varepsilon_\alpha}^\beta$, with
${\varepsilon_\alpha}^\beta$ given by (\ref{gj}). As mentioned in
\S\ref{n>1,r=2}, for $r=2$ the warp factor appearing in the internal metric
is always proportional to $\textrm{F}^2$, so that by a proper rescaling of
the internal coordinate we can set $\lambda=\tau\;\textrm{F}^2$, with
$\tau=\pm$. Equations (\ref{RiemannABCD n=1}), (\ref{kinkEq n=1a}) and the
dimensional reduced (\ref{traceless}) take then the form
\begin{subequations}\begin{eqnarray}
&&R=2k+3\sigma\varphi^2+\tau\frac{3l^2}{\varphi^4},
  \label{CentrifugalKinkEq1}\\
&&\nabla^2\varphi+2k\varphi+\sigma\varphi^3
  -\tau\frac{l^2}{\varphi^3}=0,
  \label{CentrifugalKinkEq2}\\
&&\nabla_\alpha\nabla_\beta\varphi-
  \frac{1}{2}\gamma_{\alpha\beta}\nabla^2\varphi=0,
  \label{CentrifugalKinkEq3}
\end{eqnarray}\label{CentrifugalKinkEq}\end{subequations}
reproducing the gravitational kink equations of \S\ref{r=2} up to
centripetal ($\tau=+$) or centrifugal ($\tau=-$) terms
proportional to the square of the `angular momentum' $l$
\cite{Maraner&Pachos08a}. Besides reobtaining the Kaluza-Klein
solutions (\ref{KKsc=0}), (\ref{KKsc>0}), (\ref{KKsc<0}) for $r=2$, it is
interesting to follow the fate of the gravitational kink solutions
(\ref{AKinkR}), (\ref{KinkR}) for a non-vanishing $l$. Equations
(\ref{CentrifugalKinkEq}a), (\ref{CentrifugalKinkEq}b) and
(\ref{CentrifugalKinkEq}c) are solved along the lines indicated in the
appendices A and B of Ref.~\cite{Guralnik&Iorio&Jackiw&Pi03}. By thinking of
(\ref{CentrifugalKinkEq2}) as a Newtonian equation,
$\nabla^2\varphi=V'(\varphi)$, for $\sigma=-$  we choose the integration
constant in the potential in such a way that
$V(\varphi)={\left(2K+L-\varphi^2\right)^2\left(\varphi^2-L\right)}/{4\varphi^2}$.
By differentiating and comparing with (\ref{CentrifugalKinkEq2}), we then
obtain the relations $k=K+3L/4$ and $l^2=2\tau\left(K+L/2\right)^2L$ between
the old and the new
constants. %\footnote{Inversion yields $K=\frac{4\,\sqrt[3]{4}k^2 +
%{\left(16k^3+3l\left(9 l +{\sqrt{96 k^3 + 81
%l^2}}\right)\right)}^{\frac{2}{3}}}{2\sqrt[3]{2}{\left( 16 k^3 + 3
%l\,\left(9 l +{\sqrt{96 k^3 + 81
%l^2}}\right)\right)}^{\frac{1}{3}}}-k$ and $L=\frac{4}{3}(K-k)$.}
In particular, $L$ results to be positive for $\tau=+$ and
negative for $\tau=-$. For $K>0$, the integration of the
corresponding flat-space equation yields the solution
\begin{subequations}\begin{eqnarray}
&&\textrm{g}_{\alpha\beta}=\left(
\begin{array}{cc}
-\frac{2K^3\,\textrm{sech}^4\left(\sqrt{\frac{K}{2}}\,\xi^1\right)\,
    \tanh^2\left(\sqrt{\frac{K}{2}}\,\xi^1\right)}{
    2K\,\tanh^2\left(\sqrt{\frac{K}{2}}\,\xi^1\right)+L}&0\\0&1
\end{array}\right),\label{gCentrifugalKink}\\
&&\textrm{A}_\alpha=\left(\mbox{$\pm\,K\,
{\textrm{sech}^2\left(\sqrt{\frac{K}{2}}\,\xi^1\right)}$}\;,\;0\right),\label{ACentrifugalKink}\\
&&a^1_\alpha=\left(\mbox{$\pm\frac{K\sqrt{2\tau L}}{2(2K+L)
\cosh^2\left(\sqrt{\frac{K}{2}}\,\xi^1\right)-4K}$}\;,\;0\right),\label{aCentrifugalKink}
\end{eqnarray}\label{CentrifugalKink}\end{subequations}
with the corresponding centripetal/centrifugal distortion of the
kink profile
\begin{equation}
\mbox{$\varphi(\xi)=\pm\sqrt{2K\,\tanh^2\left(\sqrt{\frac{K}{2}}\,\xi^1\right)+L}$}
\label{CentrifugalKinkProfile}
\end{equation}
and the internal warp factor
\begin{equation}
\mbox{$\lambda(\xi)=-2\;\tau\,\left(2K\tanh^2\left(\sqrt{\frac{K}{2}}\,\xi^1\right)+L\right)$}.
\end{equation}
Spacetime carries a Kaluza-Klein-like structure, complicated by a nontrivial
warp factor that cannot be set to one without introducing an explicit
internal coordinate dependence in the other entries of the metric. For
$l\rightarrow0$ the constants $L$ and $K$ respectively approach $0$ and $k$,
(\ref{CentrifugalKink}a), (\ref{CentrifugalKink}b),
(\ref{CentrifugalKinkProfile}) correctly reproduce (\ref{Kink}),
(\ref{KinkProfile}), while spacetimes reduces to (\ref{AKinkR}).
 For negative values of $L$ ($\tau=-$) the external metric
$\textrm{g}_{\alpha\beta}$, together with the corresponding scalar
curvature, is singular at $\xi^1=\pm
\sqrt{\frac{2}{K}}\textrm{arctanh}\sqrt{\frac{|L|}{2K}}$, while
the gauge field $\varphi(\xi)$ only results to be defined for
$|\xi^1|\geq\sqrt{\frac{2}{K}}\textrm{arctanh}\sqrt{\frac{|L|}{2K}}$. The
effect of the centrifugal deformation is therefore that of opening a gap in
spacetime, thus dividing it in two disconnected regions. In Figure
\ref{figure} we plot the gauge field kink profile, its centrifugal and
centripetal deformations, together
with the corresponding potential $V(\varphi)$.\\
The solutions corresponding to a negative signature of the extra
Kaluza-Klein coordinate $x^d$, are once again obtained by changing
the sign of the higher dimensional metric. As it has to be
expected, in the latter (former) case, spacetime  achieves
$\textrm{dS}_3$ ($\textrm{AdS}_3$) at infinity. We find it natural
to refer to this classes of solutions as \emph{c-kink/anti-c-kink}
spacetimes and denote them by
\begin{equation}
cK^3_s(k,l)\hskip0.5cm\mbox{and}\hskip0.5cm\textrm{A}cK^3_s(k,l),
\end{equation}
with $s=1,2$, where we give in brackets the parameters labeling the
solution.

\begin{table}[h]
\begin{center}
\caption{Solutions of Grumiller-Jackiw equations.}\label{table}
\begin{tabular}[b]{ccl}
 \hline
 \emph{rank} & \emph{nullity} &  \emph{solutions}  \\
 \hline
 $r=0$  & $n=d$ &
  $\mathbb{R}^d_s(0)$,
  $\mathbb{R}P^d_s(k)$,
  $\mathbb{R}H^d_s(k)$\\
 $r\geq2$ & $n>1$ &
  $\mathbb{C}P^\textrm{r}_\textrm{s} \left(\frac{\textrm{F}^2}{r}\right)
  \times \mathbb{R}H^n_s\left(-\frac{\textrm{F}^2}{4r}\right)$,
  $\mathbb{B}P^\textrm{r}\left(\frac{\textrm{F}^2}{r}\right)
  \times \mathbb{R}H^n_s\left(-\frac{\textrm{F}^2}{r}\right)$,
  \\
 &&
  $\mathbb{C}H^\textrm{r}_\textrm{s}\left(\frac{\textrm{F}^2}{r}\right)
  \times \mathbb{R}P^n_s\left(\frac{-\textrm{F}^2}{4r}\right)$,
  $\mathbb{B}H^\textrm{r}\left(\frac{\textrm{F}^2}{r}\right)
  \times \mathbb{R}P^n_s\left(-\frac{\textrm{F}^2}{4r}\right)$
  \\
 $r=2$  & $n>1$ &
  $\textrm{A}K^2_1(k)\times_{4k\tanh^2\sqrt{\frac{k}{2}}\xi^1}\mathbb{R}P^n_s\left(2k^2\right)$\\
 &&
  $\textrm{A}K^2_1(k)\times_{-4k\tanh^2\sqrt{\frac{k}{2}}\xi^1}\mathbb{R}H^n_s\left(-2k^2\right)$\\
 &&
  $K^2_1(k)\times_{4k\tanh^2\sqrt{\frac{k}{2}}\xi^1}\mathbb{R}H^n_s\left(-2k^2\right)$\\
 &&
  $K^2_1(k)\times_{-4k\tanh^2\sqrt{\frac{k}{2}}\xi^1}\mathbb{R}P^n_s\left(2k^2\right)$\\
 $r\geq2$ & $n=1$ &
  $\mathbb{C}^\textrm{r}_\textrm{s}\left(0\right)\times^{\pm
  \sqrt{\frac{|\textrm{F}^2|}{r}}J_{\alpha\beta}}\mathbb{R}$,
  $\mathbb{A}^\textrm{r}\left(0\right)\times^{\pm
  \sqrt{\frac{|\textrm{F}^2|}{r}}J_{\alpha\beta}}\mathbb{R}$, \\
 &&
  $\mathbb{C}P^\textrm{r}_\textrm{s}\left(
   \mbox{$\frac{\textrm{F}^2}{r}\pm\frac{4rl^2}{\textrm{F}^2}$}
   \right)\times^{\pm2l\sqrt{\frac{r}{|\textrm{F}^2|}}J_{\alpha\beta}}\mathbb{R}$, \\
 &&
  $\mathbb{B}P^\textrm{r}\left(
   \mbox{$\frac{\textrm{F}^2}{r}\pm\frac{4rl^2}{\textrm{F}^2}$}
   \right)\times^{\pm2l\sqrt{\frac{r}{|\textrm{F}^2|}}J_{\alpha\beta}}\mathbb{R}$,\\
 &&
  $\mathbb{C}H^\textrm{r}_\textrm{s}\left(
   \mbox{$\frac{\textrm{F}^2}{r}\pm\frac{4rl^2}{\textrm{F}^2}$}
   \right)\times^{\pm2l\sqrt{\frac{r}{|\textrm{F}^2|}}J_{\alpha\beta}}\mathbb{R}$,\\
 &&
  $\mathbb{B}H^\textrm{r}\left(
  \mbox{$\frac{\textrm{F}^2}{r}\pm\frac{4rl^2}{\textrm{F}^2}$}
  \right)\times^{\pm2l\sqrt{\frac{r}{|\textrm{F}^2|}}J_{\alpha\beta}}\mathbb{R}$\\
 $r=2$  & $n=1$ &
  $\textrm{A}cK^3_s(k,l)$, $cK^3_s(k,l)$\\
 $r=2\textrm{d}$ & $n=0$ &
  $\mathbb{C}P^\textrm{d}_\textrm{s}\left(\frac{F^2}{d}\right)$,
  $\mathbb{B}P^\textrm{d}\left(\frac{F^2}{d}\right)$,
  $\mathbb{C}H^\textrm{d}_\textrm{s}\left(\frac{F^2}{d}\right)$,
  $\mathbb{B}H^\textrm{d}\left(\frac{F^2}{d}\right)$\\
\hline
\end{tabular}
\end{center}
\end{table}

\section{Conclusions}
\label{conclusions}
We have shown that the equations describing the vanishing of the Weyl
conformal tensor in $d+1$-dimensional Kaluza-Klein theories, resembling
equations of motion of some $d$-dimensional Einstein-Maxwell-like theory,
admit highly symmetric solutions with maximally compatible metric and
electromagnetic structures.
Null and maximal rank solutions are respectively
real and complex/para-complex space forms. Intermediate rank solutions are
direct products of real space forms and complex/para-complex space forms
with related sectional and holomorphic/para-holomorphic sectional
curvatures. Remarkable exceptions are found for nullity-one and rank-two
gauge structures. In the former case, solutions are themselves Kaluza-Klein
spaces, with metric of a complex/para-complex space form and gauge field
proportional to the corresponding complex/para-complex structure. In the
latter, the theory supports two dimensional gravitational kinks, mixing with
the remaining dimensions through warped products and Kaluza-Klein like
structures. The covariant methods developed in Ref.~\cite{Maraner&Pachos08}
have proven extremely fruitful in obtaining intermediate rank solutions. A
summary of our results is presented in Table~\ref{table}.

\section*{Acknowledgments}
It is a pleasure to thank Roman Jackiw and Daniel Grumiller for bringing to our
attention their work on the Kaluza-Klein reduction of conformal tensors and
for related discussions.

\end{document}